\documentclass[12pt]{elsart}
\usepackage{natbib}
\usepackage[cp1251]{inputenc}
\usepackage[english]{babel}
\usepackage{color}
\usepackage[pdftex]{graphicx}

\begin{document}

\begin{frontmatter}
\title{
Flat-ended rebound indentation test for assessing viability of articular cartilage:
Application of the viscoelastic layer model}
\author{I.~Argatov},
\ead{iva1@aber.ac.uk}
\corauth[cor]{Corresponding author.}
\author{G.~Mishuris\corauthref{cor}}
\ead{ggm@aber.ac.uk}
\address{Institute of Mathematics and Physics, Aberystwyth University,
Ceredigion SY23 3BZ, Wales, UK}

\begin{abstract}
The rebound indentation test consisting of the displacement-controlled and load-controlled stages is considered for a frictionless cylindrical flat-ended indenter.
The mechanical behavior of an articular cartilage layer sample is modeled in the framework of viscoelastic model with time-independent Poisson's ratio. Closed-form analytical expressions for the contact force (in the displacement controlled stage) and for the indentation displacement (in the load-controlled stage) are presented for an arbitrary viscoelastic solid model. The case of standard viscoelastic solid model is considered in detail.
It has been established that the rebound displacement (that is the indentation displacement in the load-controlled stage) does not depend on the relaxed elastic modulus and Poisson's ratio as well as on the layer's thickness.

\end{abstract}

\end{frontmatter}

\setcounter{equation}{0}

\section{Introduction}
\label{1JtSectionI}

Indentation tests have been used for decades to identify mechanical properties of articular cartilage
\citep{Hayes_et_al_1972} and to assess its viability \citep{BroomFlachsmann2003}.
There is a wide variety of indentations tests and their possible applications. First of all, indentation tests can be classified with respect to the type of indenter used (cylindrical flat-ended, spherical, etc.). In the simplest case of cylindrical indenter, the contact area is supposed to remain constant during the indentation process, and only two mechanical variables, namely, the contact force, $P(t)$, and the indenter displacement, $w(t)$, can be recorded as functions of time  during the depthsensing indentation.
Second, indentation tests can be viewed according to the main mechanical parameter determined
(e.\,g., short-time bulk modulus \citep{HoriMockros1976}, indentation stiffness \citep{Lyyra_et_al_1999},
pulsatile dynamic stiffness \citep{Scandiucci_et_al_2006})
that, of course, depends on the mathematical model employed. For instance, the derivative $dP/dw$ is known as the indentation stiffness, and will be a variable quantity for time-dependent materials like articular cartilage. Third, indentation tests can be classified according to protocol types (e.\,g.,
multiple ramp-and-hold protocol for creep test \citep{Oyen2005}). This means that the time-history of loading is important. Finally, depending on the primary variable parameter chosen, one can distinguish displacement-controlled or load-controlled indentation tests, when $w(t)$ or $P(t)$ is assumed to be controlled, respectively. In particular, linear and non-linear monotonic displacement-controlled and force-controlled indentation loading paths were considered recently by \citet{ChengYang2009}.

In their recent paper, \citet{Brown_et_al_2009} employed a hybrid-type test consisting of two stages, one of which is displacement-controlled, while the other can be regarded as load-controlled.
In the present note, we apply the viscoelastic layer model for evaluating the so-called rebound indentation test and clarifying its applicability for assessing viability of articular cartilage.

\section{Rebound indentation test}
\label{1JtSection2}

We consider a cylindrical, flat-ended indentation test, which is composed of two stages. In the first stage, called the indentation phase, the sample is subject to loading at a constant speed $v_0$ to achieve an indentation depth of $w_0$. That is the indenter displacement is assumed to be specified according to the law
\begin{equation}
w^{(1)}(t)=v_0 t, \quad 0\leq t <t_m.
\label{1Jt(2.1)}
\end{equation}
The maximum indenter displacement at the end of the first stage is given by
\begin{equation}
w_0=v_0 t_m.
\label{1Jt(2.2)}
\end{equation}
Hence, specifying the values of $v_0$ and $w_0$, we obtain the duration of the first stage $t_m=w_0/v_0$.
Note that instead of an imposed maximum displacement, $w_0$, \citet{Brown_et_al_2009} operated with a prescribed value of maximum strain (deformation) $w_0/h$, which depends on the layer sample thickness $h$.

Further, we assume that at the indentation depth $w_0$ the load is immediately removed and the second stage, called the recovery phase, lasts for a theoretically indefinite time. In the recovery phase, we have
\begin{equation}
P^{(2)}(t)=0, \quad t\geq t_m.
\label{1Jt(2.3)}
\end{equation}

We distinguish in the notation the displacements, $w^{(1)}(t)$ and $w^{(2)}(t)$, and the contact loads, $P^{(1)}(t)$ and $P^{(2)}(t)$, corresponding to aforementioned two test stages. In the first stage, the function $w^{(1)}(t)$ is specified by Eq.~(\ref{1Jt(2.1)}), while $P^{(1)}(t)$ is unknown. On the contrary, the displacement function $w^{(2)}(t)$ is unknown, whereas the contact load $P^{(2)}(t)$  is specified by Eq.~(\ref{1Jt(2.3)}), in the second stage.

\section{Evaluation of the rebound test using a viscoelastic model}
\label{1JtSection4}

Let us assume that an articular cartilage sample may be modeled as a viscoelastic layer of thickness $h$ bonded to a rigid substrate. For the sake of simplicity, we neglect friction and assume that Poisson's ratio, $\nu$, of the layer material is time independent. These are standard assumptions made in most of the mathematical models. Then, based on the solution to the elastic contact problem of frictionless indentation of an elastic layer \citep{LebedevUfliand1958,Hayes_et_al_1972} and applying the elastic-viscoelastic correspondence principle \citep{LeeRadok1960}, one can arrive at the following equation between the indenter  displacement $w(t)$ and the applied force $P(t)$:
\begin{equation}
P(t)=\frac{2a}{1-\nu^2}\,\kappa(\alpha,\nu)
\int\limits_{0-}^t E(t-\tau)\frac{dw}{d\tau}(\tau)\,d\tau.
\label{1Jt(1.3)}
\end{equation}
Here,  $a$ is the contact radius, $\kappa(\alpha,\nu)$ is a dimensionless factor accounting the thickness effect through the ratio $\alpha=a/h$
\citep{Hayes_et_al_1972}, $E(t)$ is the relaxation modulus, $t$ is the time variable, $t=0-$ is the time moment just preceding the initial moment of contact. It has been tacitly assumed that $w(t)=0$ for $t<0$. Note that Eq.~(\ref{1Jt(1.3)}) was previously used in a number of studies \citep{Zhang_et_al_2004,Cao_et_al_2009}.

Let $E_\infty$ be the relaxed elastic modulus, that is the limit of modulus $E(t)$ at $t\to \infty$. Taking into account the relation $E(t)=E_\infty\Psi(t)$, where $\Psi(t)$ is the relaxation function, we rewrite Eq.~(\ref{1Jt(1.3)}) in the following form:
\begin{equation}
P(t)=\frac{2aE_\infty}{1-\nu^2}\,\kappa(\alpha,\nu)
\int\limits_{0-}^t \Psi(t-\tau)\frac{dw}{d\tau}(\tau)\,d\tau.
\label{1Jt(1.6)}
\end{equation}
Considering Eq.~(\ref{1Jt(1.6)}) as an integral equation with respect to $w(t)$, one arrives at the inverse relationship
\begin{equation}
w(t)=\frac{1-\nu^2}{2aE_\infty}\,\frac{1}{\kappa(\alpha,\nu)}
\int\limits_{0-}^t \Phi(t-\tau)\frac{dP}{d\tau}(\tau)\,d\tau,
\label{1Jt(1.7)}
\end{equation}
where $\Phi(t)$ is the creep function. With the appropriate relaxation function $\Psi(t)$ and creep function $\Phi(t)$, Eqs.~(\ref{1Jt(1.6)}) and (\ref{1Jt(1.7)}) can be used in the general case of viscoelastic solid.

In the indentation phase, according to Eqs.~(\ref{1Jt(2.1)}) and (\ref{1Jt(1.6)}), we will have
\begin{equation}
P^{(1)}(t)=\frac{2aE_\infty}{1-\nu^2}\,\kappa(\alpha,\nu)v_0
\int\limits_{0}^t \Psi(t-\tau)\,d\tau,\quad 0\leq t<t_m.
\label{1Jt(4.1)}
\end{equation}
The maximum contact load $P_m=P^{(1)}(t_m)$ is given by
\begin{equation}
P_m=\frac{2aE_\infty}{1-\nu^2}\,\kappa(\alpha,\nu)v_0
\int\limits_{0}^{t_m} \Psi(\tau)\,d\tau.
\label{1Jt(4.1a)}
\end{equation}

In the recovery phase, in accordance with Eq.~(\ref{1Jt(1.7)}), we can write
\begin{equation}
w^{(2)}(t) = \frac{1-\nu^2}{2aE_\infty}\,\frac{1}{\kappa(\alpha,\nu)}\Biggl\{
\int\limits_{0-}^{t_m-} \Phi(t-\tau)\frac{dP^{(1)}}{d\tau}(\tau)\,d\tau
+\int\limits_{t_m-}^{t} \Phi(t-\tau)\frac{dP^{(2)}}{d\tau}(\tau)\,d\tau\Biggr\}.
\label{1Jt(4.3)}
\end{equation}

In view of (\ref{1Jt(2.3)}), the last integral in (\ref{1Jt(4.3)}) is equal to
$-P_m\Phi(t-t_m)$, where $P_m$ is defined by (\ref{1Jt(4.1a)}). The derivative
$dP^{(1)}/d\tau$ entering the first integral in (\ref{1Jt(4.3)}) can be computed by differentiating the both sides of Eq.~(\ref{1Jt(4.1)}). Thus, taking also into account Eq.~(\ref{1Jt(4.1a)}), we can rewrite Eq.~(\ref{1Jt(4.3)}) as follows:
\begin{equation}
w^{(2)}(t) = v_0\int\limits_{0}^{t_m} [\Phi(t-\tau)-\Phi(t-t_m)] \Psi(\tau)\,d\tau.
\label{1Jt(4.4)}
\end{equation}

It is interesting and very important to observe that, in view of (\ref{1Jt(4.4)}), the rebound displacement  $w^{(2)}(t)$ does not depend on the layer material constants $E_\infty$ and $\nu$ as well as on the layer's thickness $h$.

\section{Example. Standard viscoelastic solid model}
\label{1JtSection7}

If the layer's material follows a linear standard three-parameter viscoelastic solid model, we will have
\begin{equation}
E(t)=E_\infty\bigl\{1-(1-1/\rho)\exp\bigl(-t/(\rho\tau_s)\bigr)\bigr\},
\label{1Jt(1.4)}
\end{equation}
where $\tau_s$ is the characteristic relaxation time of strain under applied step of stress, $\rho$ is the ratio of the relaxed elastic modulus $E_\infty$ (the limit of modulus $E(t)$ at $t\to \infty$) to
the unrelaxed elastic modulus $E_0$ (modulus $E(t)$ at $t=0$), i.\,e., $\rho=E_\infty/E_0<1$.

According to Eq.~(\ref{1Jt(1.4)}), we have
\begin{equation}
\Psi(t)=1-(1-1/\rho)\exp\bigl(-t/(\rho\tau_s)\bigr),\quad
\Phi(t)=1-(1-\rho)\exp(-t/\tau_s).
\label{1Jt(1.5)}
\end{equation}

After the substitution of the expressions (\ref{1Jt(1.5)}), Eq.~(\ref{1Jt(4.4)}) yields
\begin{equation}
w^{(2)}(t) = v_0 (1-\rho) \exp\Bigl(-\frac{t-t_m}{\tau_s}\Bigr)
\biggl\{t_m-\rho\tau_s+\rho\tau_s\exp\Bigl(-\frac{t_m}{\rho \tau_s}\Bigr)
\biggr\}.
\label{1Jt(4.5a)}
\end{equation}

It is interesting and important to point out that in view of (\ref{1Jt(4.5a)}) the rebound displacement  $w^{(2)}(t)$ depends on the time variable $t$ only through the exponent factor $\exp(-(t-t_m)/\tau_s)$.

\section{Rebound strain. Comparison with experimental observations}
\label{1JtSection8}

Following \citet{Brown_et_al_2009}, we determine the rebound strain, $\varepsilon_{\rm r}(t)$, as the relative elastic reverse displacement that the viscoelastic layer recoves to from the moment of unloading. In other words, we put
\begin{equation}
\varepsilon_{\rm r}(t)=\frac{w^{(1)}(t_m)-w^{(2)}(t)}{h}, \quad t\geq t_m.
\label{1Jt(3.1)}
\end{equation}
Now, taking into account Eqs.~(\ref{1Jt(2.1)}) and (\ref{1Jt(2.2)}), we rewrite Eq.~(\ref{1Jt(3.1)}) in the form
\begin{equation}
\varepsilon_{\rm r}(t)=n_0-\frac{w^{(2)}(t)}{h},
\label{1Jt(3.2)}
\end{equation}
where $n_0=w_0/h$ is the maximum strain at the end of the loading stage, and it is assumed that $t\geq t_m$.

\begin{figure}[h!]
    \centering
\vbox{
    \includegraphics [scale=0.5]{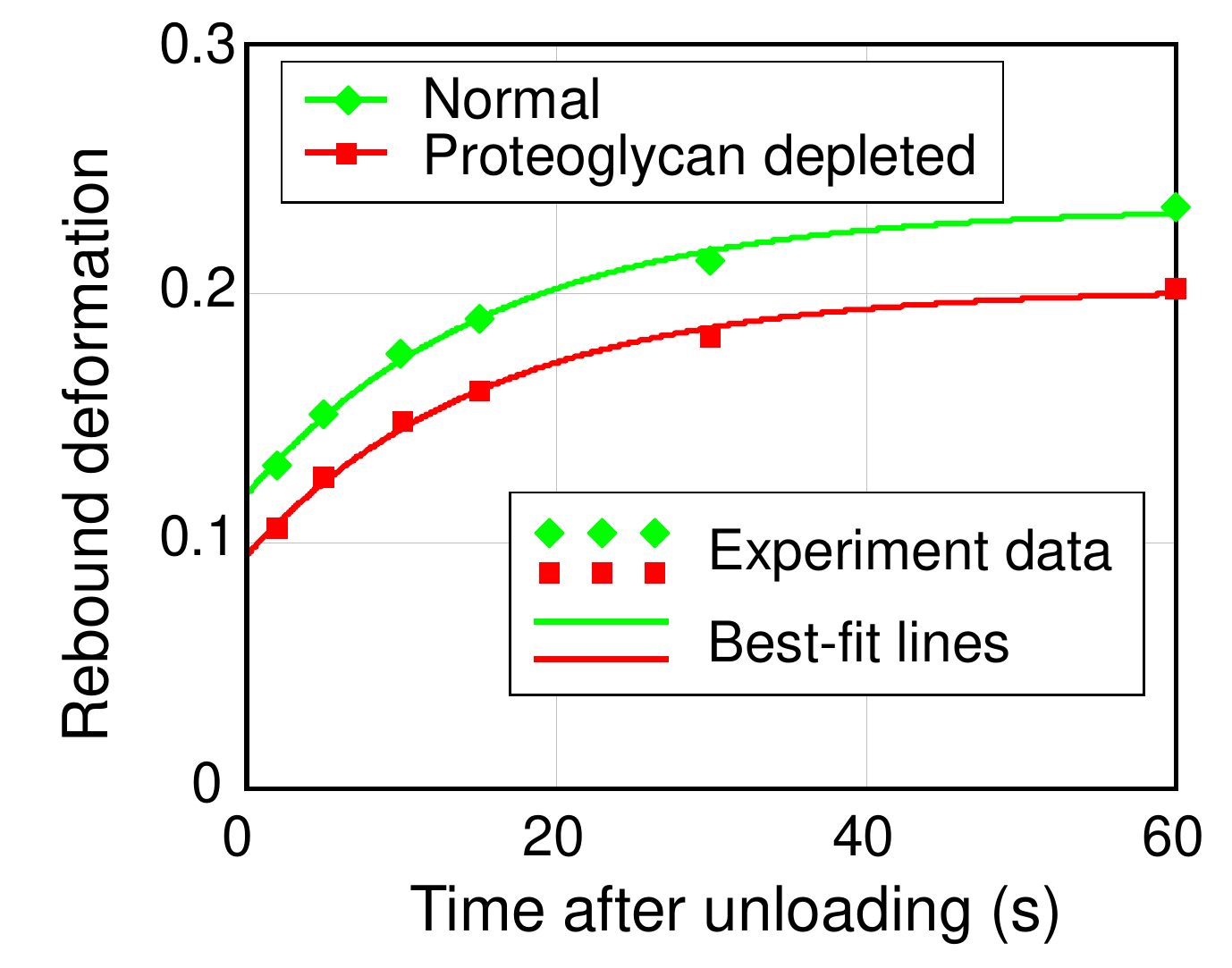}
    \caption{Rebound $\varepsilon_{\rm r}$ versus time behavior before and after one hour trypsin treatment following $0{.}1~{\rm s}^{-1}$ indentation. Experimental data were obtained by \citet{Brown_et_al_2009}. }
   }
    \label{figureR2.pdf}
\end{figure}

According to (\ref{1Jt(3.2)}) and (\ref{1Jt(4.5a)}), in the case of standard viscoelastic solid model, the rebound strain can be represented as
\begin{equation}
\varepsilon_{\rm r}(t)=n_0-K_1\exp(-(t-t_m)/\tau_s),
\label{1Jt(3.2z2)}
\end{equation}
where $K_1$ is a dimensionless constant, which is proportional to the factor $w_0/h$ and depends on the ratios $\rho$ and $\tau_s/t_m$.

The experimental points depicted in Fig.~\ref{figureR2.pdf} were obtained by \citet{Brown_et_al_2009} in the indentation tests with a relative large maximum indentation strain of 30 per cent. Based on (\ref{1Jt(3.2z2)}), we make use of the following formula
\begin{equation}
\varepsilon_{\rm r}(t)=K_2-K_1\exp(-(t-t_m)/\tau_s)
\label{1Jt(3.2z3)}
\end{equation}
with three fitting parameters $K_1$, $K_2$, and $\tau_s$. The results of the least-squares best fit of the analytical expression (\ref{1Jt(3.2z3)}) to the experimental data are as follows:
$K_1=0{.}116$, $\tau_s=15{.}87$~s, $K_2=0{.}234$ (normal cartilage) and
$K_1=0{.}107$, $\tau_s=15{.}65$~s, $K_2=0{.}202$ (artificially degraded cartilage).
It should be noted that in the both cases, the values of parameter $K_2$ are different from $0{.}3$ (which is the experimentally imposed maximum strain $n_0$).
In other words, the cartilage samples do not show complete recovery. Further, the characteristic relaxation time $\tau_s$ is found to decrease with degradation. Such a tendency is in agreement with that obtained by \citet{HayesMockros1971} in the shear creep test for human articular cartilage.

\section{Conclusion}
\label{1JtSectionC}

In the present study, the rebound indentation test has been evaluated in the framework of viscoelastic layer model. For the contact force $P^{(1)}(t)$ in the displacement controlled stage and for the indentation displacement $w^{(2)}(t)$ in the load-controlled stage, the closed-form analytical expressions (\ref{1Jt(4.1)}) and (\ref{1Jt(4.4)}), respectively, have been derived for an arbitrary viscoelastic solid model.

In the case of standard viscoelastic solid model, a simple exponential behavior was established for the rebound displacement $w^{(2)}(t)$ (see Eq.~(\ref{1Jt(4.5a)})). The theoretically predicted dependence of $w^{(2)}(t)$ on the time variable $t$ only through the exponent $\exp(-(t-t_m)/\tau_s)$ was verified using the three-parameter fitting formula for the rebound strain (\ref{1Jt(3.2z3)}).

From the view point of assessing viability of articular cartilage, the main finding of the present study is that the rebound displacement  $w^{(2)}(t)$ does not depend on the layer material constants $E_\infty$ and $\nu$ as well as on the layer's thickness $h$. At the same time, the rebound deformation mainly depends on the characteristic relaxation time $\tau_s$, which is attributed to the articular cartilage permeability.

\section*{Acknowledgements}
The authors would like to thank Dr.~M.~Stoffel (RWTH Aachen University) for bringing their attention to the paper of \citet{Brown_et_al_2009}. 
One of the authors (I.A.) also gratefully acknowledges the support from the European Union Seventh Framework Programme under contract number PIIF-GA-2009-253055.
Preliminary findings of this study were presented at a research seminar of the Laboratory for Biomechanics and Biocalometry (University of Basel). The authors are thankful to Professor A.U.~Dan Daniels and Dr.~D.~Wirz for valuable discussion.

\end{document}